\documentclass[aip,jcp,reprint,noshowkeys,superscriptaddress]{revtex4-1}
\usepackage{graphicx,dcolumn,bm,xcolor,microtype,multirow,amscd,amsmath,amssymb,amsfonts,physics,longtable,wrapfig,txfonts}
\usepackage[version=4]{mhchem}
\usepackage{natbib}
\usepackage[utf8]{inputenc}
\usepackage[T1]{fontenc}
\usepackage{txfonts}
\usepackage{soul}
\usepackage{siunitx}
\DeclareSIUnit[number-unit-product = {\,}]
\cal{cal}
\DeclareSIUnit\kcal{\kilo\cal}
\newcommand{\kcalmol}{\si{\kcal\per\mole}}

\usepackage[
	colorlinks=true,
    citecolor=blue,
    breaklinks=true
	]{hyperref}
\newcommand{\ie}{\textit{i.e.}}
\newcommand{\eg}{\textit{e.g.}}
\newcommand{\alert}[1]{\textcolor{black}{#1}}
\usepackage[normalem]{ulem}

\newcommand{\mc}{\multicolumn}
\newcommand{\fnm}{\footnotemark}
\newcommand{\fnt}{\footnotetext}

\newcommand{\SupInf}{\textcolor{blue}{supporting information}}

\newcommand{\Dtwo}{$D_{2h}$}
\newcommand{\Dfour}{$D_{4h}$}

\newcommand{\oneAg}{$1{}^1A_g$}
\newcommand{\tBoneg}{$1{}^3B_{1g}$}
\newcommand{\sBoneg}{$1{}^1B_{1g}$}
\newcommand{\twoAg}{$2{}^1A_g$}

\newcommand{\Atwog}{$1{}^3A_{2g}$}
\newcommand{\Aoneg}{$1{}^1A_{1g}$}
\newcommand{\Btwog}{$1{}^1B_{2g}$}

\begin{document}	

\newcommand{\LCPQ}{Laboratoire de Chimie et Physique Quantiques (UMR 5626), Universit\'e de Toulouse, CNRS, UPS, France}
\newcommand{\CEISAM}{Nantes Universit\'e, CNRS,  CEISAM UMR 6230, F-44000 Nantes, France}

\title{Reference Energies for Cyclobutadiene: Automerization and Excited States}

\author{Enzo \surname{Monino}*}
	\email{emonino@irsamc.ups-tlse.fr}
	\affiliation{\LCPQ}	 
\author{Martial \surname{Boggio-Pasqua}}
	\affiliation{\LCPQ}
\author{Anthony \surname{Scemama}}
	\affiliation{\LCPQ}
\author{Denis \surname{Jacquemin}}
	\affiliation{\CEISAM}
\author{Pierre-Fran\c{c}ois \surname{Loos}*}
	\email{loos@irsamc.ups-tlse.fr}
	\affiliation{\LCPQ}

\begin{abstract}
\paragraph*{Abstract:}	
Cyclobutadiene is a well-known playground for theoretical chemists and is particularly suitable to test ground- and excited-state methods.
Indeed, due to its high spatial symmetry, especially at the $D_{4h}$ square geometry but also in the $D_{2h}$ rectangular arrangement, the ground and excited states of cyclobutadiene exhibit multi-configurational characters and single-reference methods, such as \alert{standard} adiabatic time-dependent density-functional theory (TD-DFT) or \alert{standard} equation-of-motion coupled cluster (EOM-CC), are notoriously known to struggle in such situations.
In this work, using a large panel of methods and basis sets, we provide an extensive computational study of the automerization barrier (defined as the difference between the square and rectangular ground-state energies) and the vertical excitation energies at $D_{2h}$ and $D_{4h}$ equilibrium structures. 
In particular, selected configuration interaction (SCI), multi-reference perturbation theory (CASSCF, CASPT2, and NEVPT2), and coupled-cluster (CCSD, CC3, CCSDT, CC4, and CCSDTQ) calculations are performed. 
The spin-flip formalism, which is known to provide a qualitatively correct description of these diradical states, is also tested within TD-DFT (combined with numerous exchange-correlation functionals) and the algebraic diagrammatic construction [ADC(2)-s, ADC(2)-x, and ADC(3)] schemes. 
A theoretical best estimate is defined for the automerization barrier and for each vertical transition energy.
\bigskip
\begin{center}
        \boxed{\includegraphics[width=0.4\linewidth]{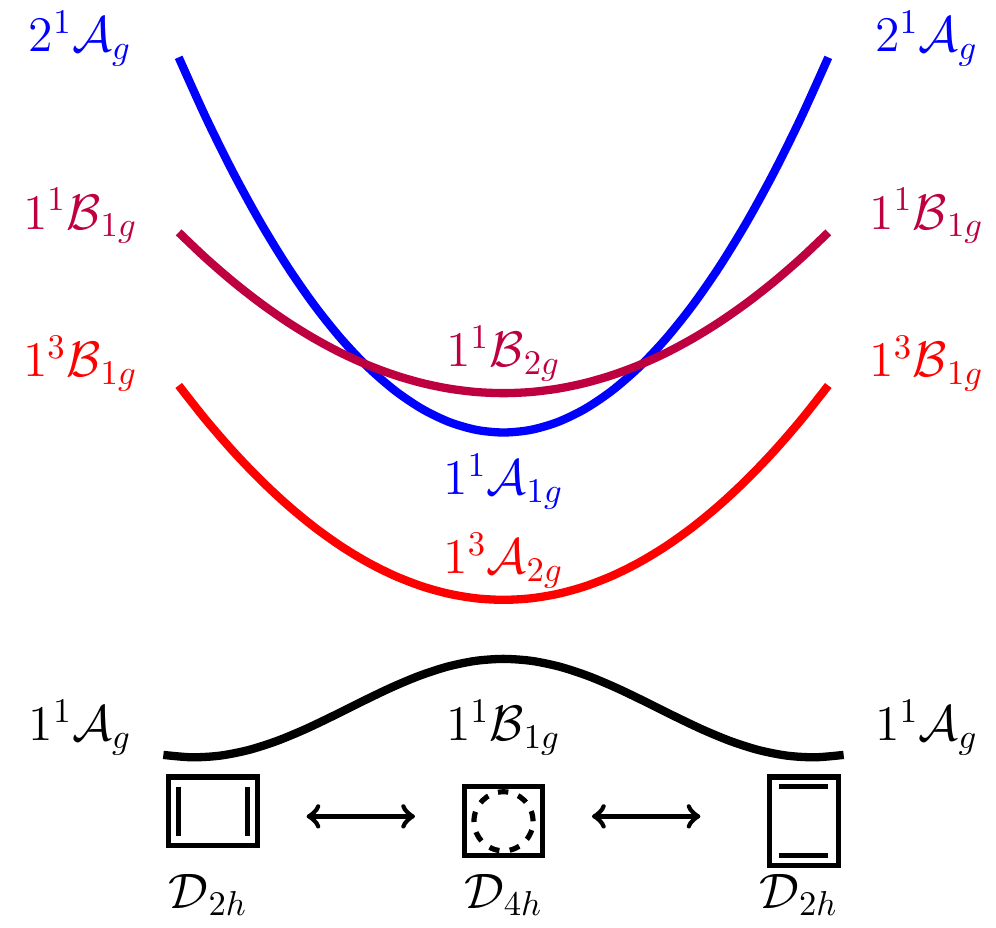}}
\end{center}
\bigskip
\end{abstract}

\maketitle

\section{Introduction}										
\label{sec:intro}

Despite the fact that excited states are involved in ubiquitous processes such as photochemistry, \cite{Bernardi_1990,Bernardi_1996,Boggio-Pasqua_2007,Klessinger_1995,Olivucci_2010,Robb_2007,VanderLugt_1969} catalysis, \cite{Kancherla_2019} and solar cells, \cite{Delgado_2010} none of the currently existing methods has shown to provide accurate excitation energies in all scenarios due to the complexity of the process, the size of the systems, the impact of the environment, and many other  factors.
Indeed, each computational model has its own theoretical and/or technical limitations and the number of possible chemical scenarios is so vast that the design of new excited-state methodologies remains a very active field of theoretical quantum chemistry.\cite{Roos_1996,Piecuch_2002b,Dreuw_2005,Krylov_2006,Sneskov_2012,Gonzales_2012,Laurent_2013,Adamo_2013,Dreuw_2015,Ghosh_2018,Blase_2020,Loos_2020a,Hait_2021,Zobel_2021}

Speaking of difficult tasks, the cyclobutadiene (CBD) molecule has been a real challenge for both experimental and theoretical chemistry for many decades. \cite{Bally_1980} 
Due to its antiaromaticity \cite{Minkin_1994} and large angular strain, \cite{Baeyer_1885} CBD presents a high reactivity making its synthesis a particularly difficult exercise. 
In the {\Dfour} symmetry, the simple H\"uckel molecular orbital theory wrongly predicts a triplet ground state (Hund's rule) with two singly-occupied frontier orbitals that are degenerate by symmetry, while state-of-the-art \textit{ab initio} methods correctly predict an open-shell singlet ground state.
This degeneracy is lifted by the so-called pseudo Jahn-Teller effect, \ie, by a descent in symmetry (from {\Dfour} to {\Dtwo} point group) via a geometrical distortion of the molecule, leading to a closed-shell singlet ground state in the rectangular geometry (see below). 
This was confirmed by several experimental studies by Pettis and co-workers \cite{Reeves_1969} and others. \cite{Irngartinger_1983,Ermer_1983,Kreile_1986}

\begin{figure}
	\includegraphics[width=\linewidth]{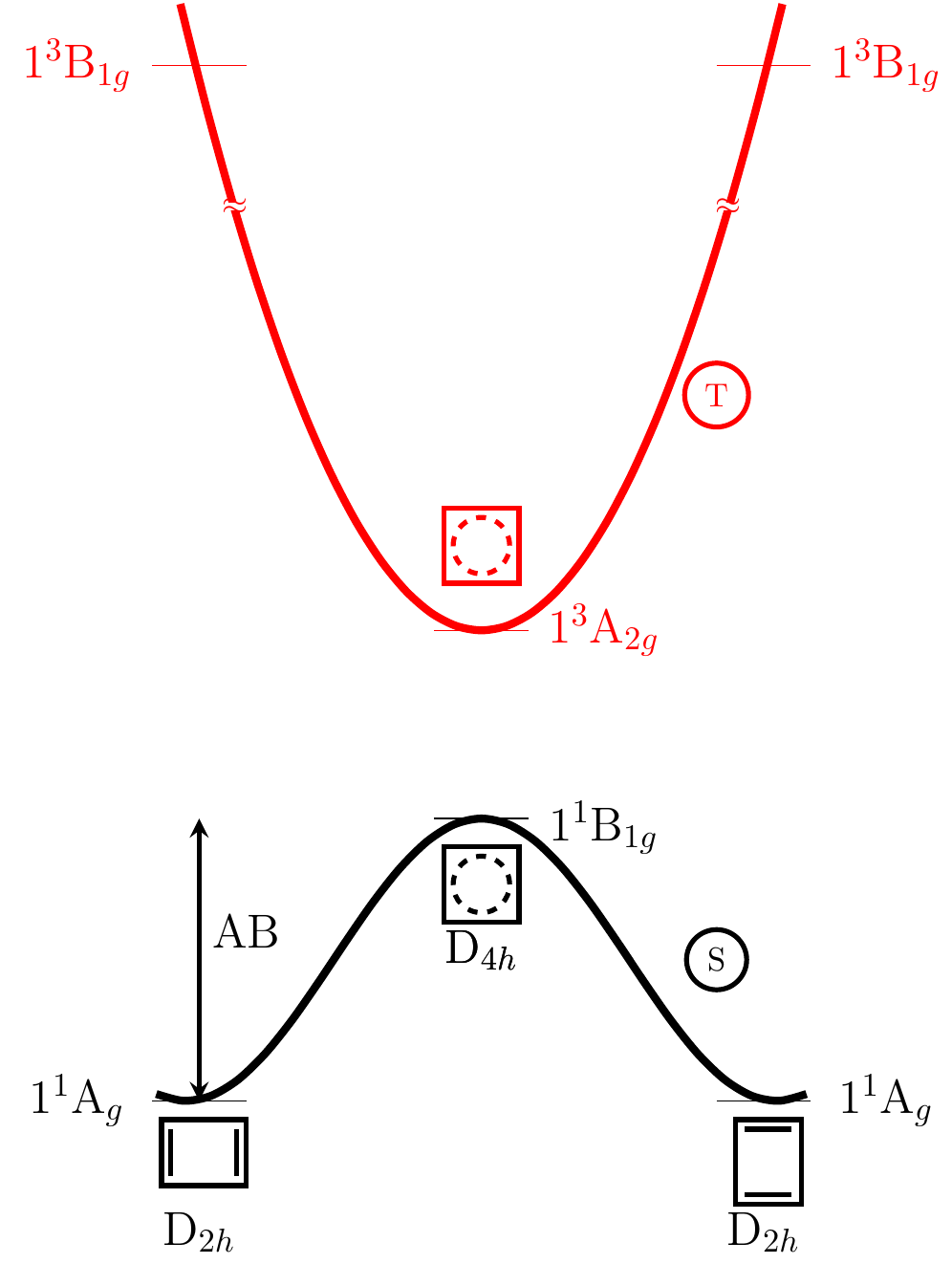}
	\caption{Pictorial representation of the ground and lowest excited states of CBD and the properties under investigation. 
	The singlet ground state (S) and triplet (T) properties are colored in black and red, respectively.
	The automerization barrier (AB) is also represented.}
\label{fig:CBD}
\end{figure}

In the {\Dtwo} symmetry, the {\oneAg} ground state has a weak multi-configurational character with well-separated frontier orbitals that can be described by single-reference methods. 
However, in the {\Dfour} symmetry, the {\sBoneg} ground state is a diradical that has two degenerate singly occupied frontier orbitals.
Therefore, one must take into account, at least, two electronic configurations to properly model this multi-configurational scenario. 
Of course, \alert{standard} single-reference methods are naturally unable to describe such situations. 
Interestingly, the {\sBoneg} ground state of the square arrangement is a transition state in the automerization reaction between the two rectangular structures (see Fig.~\ref{fig:CBD}), while the lowest triplet state,  {\Atwog}, is a minimum on the triplet potential energy surface in the {\Dfour} arrangement. 
The automerization barrier (AB) is thus defined as the difference between the square and rectangular ground-state energies.
The energy of this barrier is estimated, experimentally, in the range of \SIrange{1.6}{10}{\kcalmol}, \cite{Whitman_1982} while previous state-of-the-art \textit{ab initio} calculations yield values in the \SIrange{7}{9}{\kcalmol} range. \cite{Eckert-Maksic_2006,Li_2009,Shen_2012,Zhang_2019}

The lowest-energy excited states of CBD in both symmetries are represented in Fig.~\ref{fig:CBD}, where we have reported the {\oneAg} and {\tBoneg} states for the rectangular geometry and the {\sBoneg} and {\Atwog} states for the square one. 
Due to the energy scale, the higher-energy states ({\sBoneg} and {\twoAg} for {\Dtwo} and {\Aoneg} and {\Btwog} for {\Dfour}) are not shown. 
Interestingly, the {\twoAg} and {\Aoneg} states have a strong contribution from doubly-excited configurations and these so-called double excitations \cite{Loos_2019} are known to be inaccessible with \alert{standard} adiabatic time-dependent density-functional theory (TD-DFT) \cite{Runge_1984,Casida_1995,Tozer_2000,Maitra_2004,Cave_2004,Levine_2006,Elliott_2011,Maitra_2012,Maitra_2017} and \alert{remain challenging for standard hierarchy of EOM-CC methods that are using ground-state Hartree-Fock reference}. \cite{Kucharski_1991,Kallay_2004,Hirata_2000,Hirata_2004}

In order to tackle the problem of multi-configurational character and double excitations, we have explored several approaches. 
The most evident way is to rely on \alert{multi-reference} methods, which are naturally designed to address such scenarios.
Among these methods, one can mention the complete-active-space self-consistent field (CASSCF) method, \cite{Roos_1996} its second-order perturbatively-corrected variant (CASPT2) \cite{Andersson_1990,Andersson_1992,Roos_1995a} and the second-order $n$-electron valence state perturbation theory (NEVPT2) formalism. \cite{Angeli_2001,Angeli_2001a,Angeli_2002}

Another way to deal with double excitations and multi-reference situations is to use high level truncation of the EOM formalism \cite{Rowe_1968,Stanton_1993} of CC theory. \cite{Kucharski_1991,Kallay_2003,Kallay_2004,Hirata_2000,Hirata_2004}
However, to provide a correct description of these situations, one has to take into account, at the very least, contributions from the triple excitations in the CC expansion. \cite{Watson_2012,Loos_2018a,Loos_2019,Loos_2020b}
Although multi-reference CC methods have been designed, \cite{Jeziorski_1981,Mahapatra_1998,Mahapatra_1999,Lyakh_2012,Kohn_2013} they are computationally demanding and remain far from being black-box.

In this context, an interesting alternative to \alert{multi-reference} and CC methods is provided by selected configuration interaction (SCI) methods, \cite{Bender_1969,Whitten_1969,Huron_1973,Giner_2013,Evangelista_2014,Giner_2015,Caffarel_2016b,Holmes_2016,Tubman_2016,Liu_2016,Ohtsuka_2017,Zimmerman_2017,Coe_2018,Garniron_2018} which are able to provide near full CI (FCI) ground- and excited-state energies of small molecules. \cite{Caffarel_2014,Caffarel_2016a,Scemama_2016,Holmes_2017,Li_2018,Scemama_2018,Scemama_2018b,Li_2020,Loos_2018a,Chien_2018,Loos_2019,Loos_2020b,Loos_2020c,Loos_2020e,Garniron_2019,Eriksen_2020,Yao_2020,Williams_2020,Veril_2021,Loos_2021,Damour_2021}
For example, the \textit{Configuration Interaction using a Perturbative Selection made Iteratively} (CIPSI) method limits the exponential increase of the size of the CI expansion by retaining the most energetically relevant determinants only, using a second-order energetic criterion to select perturbatively determinants in the FCI space. \cite{Huron_1973,Giner_2013,Giner_2015,Garniron_2017,Garniron_2018,Garniron_2019}
Nonetheless, SCI methods remain very expensive and can be applied to a limited number of situations.

Finally, another option to deal with these chemical scenarios is to rely on the spin-flip formalism, established by Krylov in 2001, \cite{Krylov_2001a,Krylov_2001b,Krylov_2002,Casanova_2020} where one accesses the ground and doubly-excited states via a single (spin-flip) de-excitation and excitation from the lowest triplet state, respectively.
\alert{One drawback of spin-flip methods is spin contamination} (\ie, the artificial mixing of electronic states with different spin multiplicities) due not only to the spin incompleteness in the spin-flip expansion but also to the potential spin contamination of the reference configuration. \cite{Casanova_2020} 
One can address part of this issue by increasing the excitation order or by complementing the spin-incomplete configuration set with the missing configurations. \cite{Sears_2003,Casanova_2008,Huix-Rotllant_2010,Li_2010,Li_2011a,Li_2011b,Zhang_2015,Lee_2018}
\alert{Note that one can quantify the polyradical character associated to a given electronic state using Head-Gordon's index \cite{Head-Gordon_2003} that provides a measure of the number of unpaired electrons. \cite{Orms_2018}}

In the present work, we define highly-accurate reference values and investigate the accuracy of each family of computational methods mentioned above on the automerization barrier and the low-lying excited states of CBD at the {\Dtwo} and {\Dfour} ground-state geometries. 
Computational details are reported in Sec.~\ref{sec:compmet}.
Section \ref{sec:res} is devoted to the discussion of our results. 
Finally, our conclusions are drawn in Sec.~\ref{sec:conclusion}.

\section{Computational details}
\label{sec:compmet}

\subsection{Selected configuration interaction calculations}
\label{sec:SCI}
For the SCI calculations, we rely on the CIPSI algorithm implemented in QUANTUM PACKAGE, \cite{Garniron_2019} which iteratively select determinants in the FCI space.
To treat electronic states on an equal footing, we use a state-averaged formalism where the ground and excited states are expanded with the same set of determinants but with different CI coefficients. 
Note that the determinant selection for these states are performed simultaneously via the protocol described in Refs.~\onlinecite{Scemama_2019,Garniron_2019}.

For a given size of the variational wave function and for each electronic state, the CIPSI energy is the sum of two terms: the variational energy obtained by diagonalization of the CI matrix in the reference space $E_\text{var}$ and a second-order perturbative correction $E_\text{PT2}$ which estimates the contribution of the external determinants that are not included in the variational space at a given iteration.
The sum of these two energies is, for large enough wave functions, an estimate of the FCI energy of a given state, \ie, $E_\text{FCI} \approx E_\text{var} + E_\text{PT2}$.
It is possible to estimate more precisely the FCI energy via an extrapolation procedure, where the variational energy is extrapolated to $E_\text{PT2} = 0$. \cite{Holmes_2017} 
Excitation energies are then computed as differences of extrapolated total energies. \cite{Chien_2018,Loos_2018a,Loos_2019,Loos_2020b,Loos_2020c}
Additionally, an error bar can be provided thanks to a recent method based on Gaussian random variables that is described in Ref.~\onlinecite{Veril_2021}.
This type of extrapolation procedures is now routine in SCI and similar techniques. \cite{Eriksen_2020,Loos_2020e,Eriksen_2021}

\subsection{Coupled-cluster calculations}
\label{sec:CC}
Coupled-cluster theory provides a hierarchy of methods that yields increasingly accurate ground state energies by ramping up the maximum excitation degree of the cluster operator: \cite{Cizek_1966,Paldus_1972,Crawford_2000,Piecuch_2002b,Bartlett_2007,Shavitt_2009} CC with singles and doubles (CCSD), \cite{Cizek_1966,Purvis_1982} CC with singles, doubles, and triples (CCSDT), \cite{Noga_1987a,Scuseria_1988} CC with singles, doubles, triples, and quadruples (CCSDTQ), \cite{Oliphant_1991,Kucharski_1991a,Kucharski_1992} etc.
As mentioned above, CC theory can be extended to excited states via the EOM formalism, \cite{Rowe_1968,Stanton_1993} where one diagonalizes the similarity-transformed Hamiltonian in a CI basis of excited determinants yielding the following systematically improvable family of methods for neutral excited states:\cite{Noga_1987a,Koch_1990b,Kucharski_1991,Stanton_1993,Christiansen_1998,Kucharski_2001,Kowalski_2001,Kallay_2003,Kallay_2004,Hirata_2000,Hirata_2004} EOM-CCSD, EOM-CCSDT, EOM-CCSDTQ, etc.
In the following, we will omit the prefix EOM for the sake of conciseness.
Alternatively to the ``complete'' CC models, one can also employ the CC2, \cite{Christiansen_1995,Hattig_2000} CC3, \cite{Christiansen_1995,Koch_1995} and CC4 \cite{Kallay_2005,Matthews_2020,Loos_2021} methods which can be seen as cheaper approximations of CCSD, CCSDT, and CCSDTQ by skipping the most expensive terms and avoiding the storage of high-order amplitudes. 

Here, we have performed CC calculations using various codes. 
Typically, CCSD, CCSDT, and CCSDTQ as well as CC3 and CC4 calculations are achieved with CFOUR, \cite{Matthews_2020} with which only singlet excited states can be computed (except for CCSD). 
In some cases, we have also computed (singlet and triplet) excitation energies and properties (such as the percentage of single excitations involved in a given transition, namely $\%T_1$) at the CC3 level with DALTON \cite{Aidas_2014} and at the CCSDT level with MRCC. \cite{mrcc} 

To avoid having to perform multi-reference CC calculations or high-level CC calculations in the restricted open-shell or unrestricted formalisms, it is worth mentioning that, for the {\Dfour} arrangement, we have considered the lowest \textit{closed-shell} \alert{singlet state of {$A_g$} symmetry} as reference. 
Hence, the open-shell ground state, {\sBoneg}, and the {\Btwog} state appear as a de-excitation and an excitation, respectively.
With respect to \alert{this closed-shell reference}, {\sBoneg} has a dominant double excitation character, while {\Btwog} has a dominant single excitation character, hence their contrasting convergence behaviors with respect to the order of the CC expansion (see below). 

\subsection{\alert{Multi-reference} calculations}
\label{sec:Multi}
State-averaged CASSCF (SA-CASSCF) calculations are performed for vertical transition energies, whereas state-specific CASSCF is used for computing the automerization barrier. \cite{Werner_2020} 
For each excited state, a set of state-averaged orbitals is computed by taking into account the excited state of interest as well as the ground state (even if it has a different symmetry).
Two active spaces have been considered: (i) a minimal (4e,4o) active space including the valence $\pi$ orbitals, and (ii) an extended (12e,12o) active space where we have additionally included the $\sigma_\text{CC}$ and $\sigma_\text{CC}^*$ orbitals.
For ionic excited states, like the {\sBoneg} state of CBD, it is particularly important to take into account the $\sigma$-$\pi$ coupling. \cite{Davidson_1996,Angeli_2009,BenAmor_2020}

On top of this CASSCF treatment, CASPT2 calculations are performed within the RS2 contraction scheme, while the NEVPT2 energies are computed within both the partially contracted (PC) and strongly contracted (SC) schemes. \cite{Angeli_2001,Angeli_2001a,Angeli_2002} 
Note that PC-NEVPT2 is theoretically more accurate than SC-NEVPT2 due to the larger number of external configurations and greater flexibility. 
In order to avoid the intruder state problem in CASPT2, a real-valued level shift of \SI{0.3}{\hartree} is set, \cite{Roos_1995b,Roos_1996} with an additional ionization-potential-electron-affinity (IPEA) shift of \SI{0.25}{\hartree} to avoid systematic underestimation of the vertical excitation energies. \cite{Ghigo_2004,Schapiro_2013,Zobel_2017,Sarkar_2022}
\alert{For the sake of comparison and completeness, for the (4e,4o) active space, we also report (in the {\SupInf}) multi-reference CI calculations including Davidson correction (MRCI+Q). \cite{Knowles_1988,Werner_1988}}
All these calculations are carried out with MOLPRO. \cite{Werner_2020} 

\subsection{Spin-flip calculations}
\label{sec:sf}
Within the spin-flip formalism, one considers the lowest triplet state as reference instead of the singlet ground state.
Ground-state energies are then computed as sums of the triplet reference state energy and the corresponding de-excitation energy.
Likewise, excitation energies with respect to the singlet ground state are computed as differences of excitation energies with respect to the reference triplet state.

Nowadays, spin-flip techniques are broadly accessible thanks to intensive developments in the electronic structure community (see Ref.~\onlinecite{Casanova_2020} and references therein).
Here, we explore the spin-flip version \cite{Lefrancois_2015} of the algebraic-diagrammatic construction \cite{Schirmer_1982} (ADC) using the standard and extended second-order ADC schemes, SF-ADC(2)-s \cite{Trofimov_1997,Dreuw_2015} and SF-ADC(2)-x, \cite{Dreuw_2015} as well as its third-order version, SF-ADC(3). \cite{Dreuw_2015,Trofimov_2002,Harbach_2014} 
These calculations are performed using Q-CHEM 5.4.1. \cite{qchem} 
The spin-flip version of our recently proposed composite approach, namely SF-ADC(2.5), \cite{Loos_2020d} where one simply averages the SF-ADC(2)-s and SF-ADC(3) energies, is also tested in the following.

We have also carried out spin-flip calculations within the TD-DFT framework (SF-TD-DFT), \cite{Shao_2003} with the same Q-CHEM 5.2.1 code. \cite{qchem}
The B3LYP, \cite{Becke_1988b,Lee_1988a,Becke_1993b} PBE0 \cite{Adamo_1999a,Ernzerhof_1999} and BH\&HLYP global hybrid GGA functionals are considered, which contain 20\%, 25\%, 50\% of exact exchange, respectively. 
These calculations are labeled as SF-TD-B3LYP, SF-TD-PBE0, and SF-TD-BH\&HLYP in the following.
Additionally, we have also computed SF-TD-DFT excitation energies using range-separated hybrid (RSH) functionals: CAM-B3LYP (19\% of short-range exact exchange and 65\% at long range), \cite{Yanai_2004a} LC-$\omega$PBE08 (0\% of short-range exact exchange and 100\% at long range), \cite{Weintraub_2009a} and $\omega$B97X-V (16.7\% of short-range exact exchange and 100\% at long range). \cite{Mardirossian_2014}
Finally, the hybrid meta-GGA functional M06-2X (54\% of exact exchange) \cite{Zhao_2008} and the RSH meta-GGA functional M11 (42.8\% of short-range exact exchange and 100\% at long range) \cite{Peverati_2011} are also employed.
Note that all SF-TD-DFT calculations are done within the Tamm-Dancoff approximation. \cite{Hirata_1999}

\alert{There also exist spin-flip extensions of EOM-CC methods, \cite{Krylov_2001a,Levchenko_2004,Manohar_2008,Casanova_2009a,Dutta_2013} and we consider here the spin-flip version of EOM-CCSD, named SF-EOM-CCSD. \cite{Krylov_2001a}
Additionally, Manohar and Krylov introduced a non-iterative triples correction to EOM-CCSD and extended it to the spin-flip variant. \cite{Manohar_2008} 
Two types of triples corrections were proposed: (i) EOM-CCSD(dT) that uses the diagonal elements of the similarity-transformed CCSD Hamiltonian, and (ii) EOM-CCSD(fT) where the Hartree-Fock orbital energies are considered instead.}

\subsection{Theoretical best estimates}
\label{sec:TBE}
When technically possible, each level of theory is tested with four Gaussian basis sets, namely, 6-31+G(d) and aug-cc-pVXZ with X $=$ D, T, and Q. \cite{Dunning_1989} 
This helps us to assess the convergence of each property with respect to the size of the basis set.
More importantly, for each studied quantity (i.e., the automerization barrier and the vertical excitation energies), we provide a theoretical best estimate (TBE) established in the aug-cc-pVTZ basis. 
These TBEs are defined using extrapolated CCSDTQ/aug-cc-pVTZ values except in a single occasion where the NEVPT2(12,12) value is used.

The extrapolation of the CCSDTQ/aug-cc-pVTZ values is done via a ``pyramidal'' scheme, where we employ systematically the most accurate level of theory and the largest basis set available.
The viability of this scheme lies on the transferability of basis set effects within wave function methods (see below). 
For example, when CC4/aug-cc-pVTZ and CCSDTQ/aug-cc-pVDZ data are available, we proceed via the following basis set extrapolation:
\begin{equation}
	\label{eq:aug-cc-pVTZ}
	\Delta \Tilde{E}^{\text{CCSDTQ}}_{\text{aug-cc-pVTZ}} 
	= \Delta E^{\text{CCSDTQ}}_{\text{aug-cc-pVDZ}} 
	+ \qty[ \Delta E^{\text{CC4}}_{\text{aug-cc-pVTZ}} - \Delta E^{\text{CC4}}_{\text{aug-cc-pVDZ}} ],
\end{equation}
while, when only CCSDTQ/6-31G+(d) values are available, we further extrapolate the CCSDTQ/aug-cc-pVDZ value as follows:
\begin{equation}
	\label{eq:aug-cc-pVDZ}
	\Delta \Tilde{E}^{\text{CCSDTQ}}_{\text{aug-cc-pVDZ}} 
	= \Delta E^{\text{CCSDTQ}}_{6-31\text{G}+\text{(d)}} 
	+ \qty[ \Delta E^{\text{CC4}}_{\text{aug-cc-pVDZ}} - \Delta E^{\text{CC4}}_{6-31\text{G}+\text{(d)}} ].
\end{equation}
If we lack the CC4 data, we can follow the same philosophy and rely on CCSDT (for single excitations) or NEVPT2 (for double excitations).
For example, 
\begin{equation}
	\label{eq:CC4_aug-cc-pVTZ}
	\Delta \Tilde{E}^{\text{CC4}}_{\text{aug-cc-pVTZ}} 
	= \Delta E^{\text{CC4}}_{\text{aug-cc-pVDZ}} 
	+ \qty[ \Delta E^{\text{CCSDT}}_{\text{aug-cc-pVTZ}} - \Delta E^{\text{CCSDT}}_{\text{aug-cc-pVDZ}} ], 
\end{equation}
and so on.
If neither CC4, nor CCSDT are feasible, then we rely on PC-NEVPT2(12,12). 
The procedures applied for each extrapolated value are explicitly mentioned as footnote in the tables.
Note that, due to error bar inherently linked to the CIPSI calculations (see Sec.~\ref{sec:SCI}), these are mostly used as an additional safety net to further check the convergence of the CCSDTQ estimates.

Additional tables gathering these TBEs as well as literature data for the automerization barrier and the vertical excitation energies can be found in the {\SupInf}.
 

\section{Results and discussion}
\label{sec:res}

\subsection{Geometries}
\label{sec:geometries}
Two different sets of geometries obtained with different levels of theory are considered for the automerization barrier and the excited states of the CBD molecule. 
First, because the automerization barrier is obtained as a difference of energies computed at distinct geometries, it is paramount to obtain these at the same level of theory. 
However, due to the fact that the ground state of the square arrangement is a transition state of singlet open-shell nature, it is technically difficult to optimize the geometry with high-order CC methods.
Therefore, we rely on CASPT2(12,12)/aug-cc-pVTZ for both the {\Dtwo} and {\Dfour} ground-state structures. 
(Note that these optimizations are done without IPEA shift but with a level shift and a state-specific reference CASSCF wave function.)
Second, because the vertical transition energies are computed for a particular equilibrium geometry, we can afford to use different methods for the rectangular and square structures.
Hence, we rely on CC3/aug-cc-pVTZ to compute the equilibrium geometry of the {\oneAg} state in the rectangular ({\Dtwo}) arrangement and the restricted open-shell (RO) version of CCSD(T)/aug-cc-pVTZ to obtain the equilibrium geometry of the {\Atwog} state in the square ({\Dfour}) arrangement. 
These two geometries are the lowest-energy equilibrium structure of their respective spin manifold (see Fig.~\ref{fig:CBD}).
The cartesian coordinates of these geometries are provided in the {\SupInf}.
Table \ref{tab:geometries} reports the key geometrical parameters obtained at these levels of theory as well as previous geometries computed by Manohar and Krylov at the CCSD(T)/cc-pVTZ level.
One notes globally satisfying agreement between the tested methods with variations of the order of \SI{0.01}{\angstrom} only.

\begin{squeezetable}
\begin{table}
	\caption{Optimized geometries associated with several states of CBD computed with various levels of theory. 
	Bond lengths are in \si{\angstrom} and angles ($\angle$) are in degree.}
	\label{tab:geometries}
	\begin{ruledtabular}
		\begin{tabular}{lllrrr}
			State	&	Method				&	\ce{C=C}	&  \ce{C-C}	& \ce{C-H} & $\angle\,\ce{H-C=C}$	\\
			\hline
			{\Dtwo} ({\oneAg}) &
			CASPT2(12,12)/aug-cc-pVTZ\fnm[1] & 1.354 & 1.566 & 1.077 & 134.99 \\
			&CC3/aug-cc-pVTZ\fnm[1]  &  1.344 & 1.565 & 1.076 & 135.08 \\
			&CCSD(T)/cc-pVTZ\fnm[2] & 1.343 & 1.566 & 1.074 & 135.09\\
			{\Dfour} ({\sBoneg}) &
			CASPT2(12,12)/aug-cc-pVTZ\fnm[1] & 1.449 & 1.449 & 1.076 & 135.00 \\
			{\Dfour} ({\Atwog}) &
			CASPT2(12,12)/aug-cc-pVTZ\fnm[1] & 1.445 & 1.445 & 1.076 & 135.00 \\
			&RO-CCSD(T)/aug-cc-pVTZ\fnm[1] & 1.439 & 1.439 & 1.075 & 135.00\\
			&RO-CCSD(T)/cc-pVTZ\fnm[2] & 1.439 & 1.439 & 1.073 & 135.00\\
\end{tabular}
	\end{ruledtabular}
	\fnt[1]{This work.}
	\fnt[2]{From Ref.~\onlinecite{Manohar_2008}.}
\end{table}
\end{squeezetable}

\subsection{Automerization barrier}
\label{sec:auto}

\begin{squeezetable}
\begin{table}
	\caption{Automerization barrier (in \kcalmol) of CBD computed with various computational methods and basis sets.
	The values in square parenthesis have been obtained by extrapolation via the procedure described in the corresponding footnote.
	The TBE/aug-cc-pVTZ value is highlighted in bold.}
	\label{tab:auto_standard}
	\begin{ruledtabular}
		\begin{tabular}{lrrrr}
									&	\mc{4}{c}{Basis sets}	\\
										\cline{2-5}
			Method					&	6-31+G(d)	&	aug-cc-pVDZ &	aug-cc-pVTZ	& aug-cc-pVQZ\\
			\hline
SF-TD-B3LYP & $18.59$ & $18.64$ & $19.34$ & $19.34$ \\
SF-TD-PBE0& $17.18$ & $17.19$ & $17.88$ & $17.88$ \\
SF-TD-BH\&HLYP & $11.90$ & $12.02$ & $12.72$ & $12.73$ \\
SF-TD-M06-2X & $9.32$ & $9.62$ & $10.35$ & $10.37$ \\
SF-TD-CAM-B3LYP& $18.05$ & $18.10$ & $18.83$ & $18.83$ \\
SF-TD-$\omega$B97X-V & $18.26$ & $18.24$ & $18.94$ & $18.92$ \\
SF-TD-LC-$\omega$PBE08 & $19.05$ & $18.98$ & $19.74$ & $19.71$ \\
SF-TD-M11 & $11.03$ & $10.25$ & $11.22$ & $11.12$ \\[0.1cm]
SF-ADC(2)-s & $6.69$ & $6.98$ & $8.63$ &  \\
SF-ADC(2)-x & $8.63$ & $8.96$ &$10.37$ & \\
SF-ADC(2.5) & $7.36$ & $7.76$ & $9.11$ & \\
SF-ADC(3) & $8.03$ & $8.54$ & $9.58$  \\
\alert{SF-EOM-CCSD} & \alert{$5.86$} & \alert{$6.27$} & \alert{$7.40$} \\[0.1cm]
CASSCF(4,4) & $6.17$ & $6.59$ & $7.38$ & $7.41$ \\
CASPT2(4,4) & $6.56$ & $6.87$ & $7.77$ & $7.93$ \\
SC-NEVPT2(4,4) & $7.95$ & $8.31$ & $9.23$ & $9.42$ \\
PC-NEVPT2(4,4) & $7.95$ & $8.33$ & $9.24$ & $9.41$ \\
CASSCF(12,12) & $10.19$ & $10.75$ & $11.59$ & $11.62$ \\
CASPT2(12,12) & $7.24$ & $7.53$ & $8.51$ & $8.71$ \\
SC-NEVPT2(12,12) & $7.10$ & $7.32$ & $8.29$ & $8.51$ \\
PC-NEVPT2(12,12) & $7.12$ & $7.33$ & $8.28$ & $8.49$ \\[0.1cm]
CCSD & $8.31$ & $8.80$ & $9.88$ & $10.10$ \\
CC3 & $6.59$ & $6.89$ & $7.88$ & $8.06$ \\
CCSDT & $7.26$ & $7.64$ & $8.68$ &$[ 8.86]$\fnm[1]  \\
CC4 & $7.40$ & $7.78$ & $[ 8.82]$\fnm[2] &  $[ 9.00]$\fnm[3]\\
CCSDTQ & $7.51$ & $[ 7.89]$\fnm[4]& $[\bf  8.93]$\fnm[5]& $[ 9.11]$\fnm[6]\\
		\end{tabular}
	\end{ruledtabular}
	\fnt[1]{Value obtained using CCSDT/aug-cc-pVTZ corrected by the difference between CC3/aug-cc-pVQZ and CC3/aug-cc-pVTZ.}
	\fnt[2]{Value obtained using CC4/aug-cc-pVDZ corrected by the difference between CCSDT/aug-cc-pVTZ and CCSDT/aug-cc-pVDZ.}
	\fnt[3]{Value obtained using CC4/aug-cc-pVTZ corrected by the difference between CCSDT/aug-cc-pVQZ and CCSDT/aug-cc-pVTZ.}
	\fnt[4]{Value obtained using CCSDTQ/6-31+G(d) corrected by the difference between CC4/aug-cc-pVDZ basis and CC4/6-31+G(d).}
	\fnt[5]{TBE value obtained using CCSDTQ/aug-cc-pVDZ corrected by the difference between CC4/aug-cc-pVTZ and CC4/aug-cc-pVDZ.}
	\fnt[6]{Value obtained using CCSDTQ/aug-cc-pVTZ corrected by the difference between CC4/aug-cc-pVQZ and CC4/aug-cc-pVTZ.}
\end{table}
\end{squeezetable}

\begin{figure*}
\includegraphics[width=0.8\linewidth]{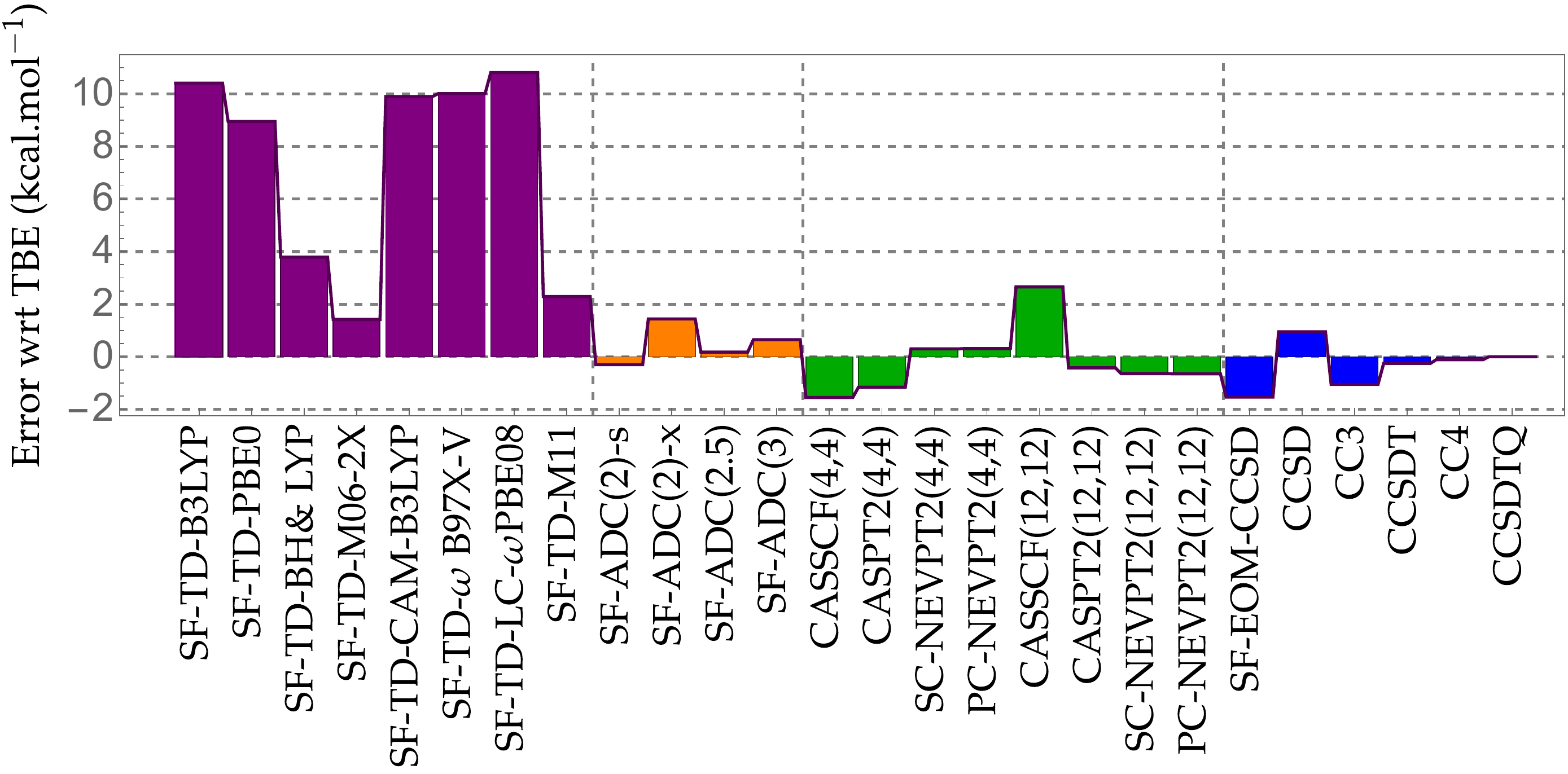}
\caption{Error (with respect to the TBE) in the automerization barrier (in \si{\kcalmol}) of CBD at various levels of theory using the aug-cc-pVTZ basis.
See {\SupInf} for the total energies.}
\label{fig:AB}
\end{figure*}

The results concerning the automerization barrier are reported in Table \ref{tab:auto_standard} for various basis sets and shown in Fig.~\ref{fig:AB} for the aug-cc-pVTZ basis.
Our TBE with this basis set is \SI{8.93}{\kcalmol}, which is in excellent agreement with previous studies \cite{Eckert-Maksic_2006,Li_2009,Shen_2012,Zhang_2019,Gururangan_2021,Deustua_2021} (see {\SupInf}).

First, one can see large variations of the energy barrier at the SF-TD-DFT level, with differences as large as \SI{10}{\kcalmol} between the different functionals for a given basis set.  
Nonetheless, it is clear that the performance of a given functional is directly linked to the amount of exact exchange at short range.
Indeed, hybrid functionals with approximately 50\% of short-range exact exchange (\eg, BH\&HLYP, M06-2X, and M11) perform significantly better than the functionals having a small fraction of short-range exact exchange (\eg, B3LYP, PBE0, CAM-B3LYP, $\omega$B97X-V, and LC-$\omega$PBE08).
However, they are still off by \SIrange{1}{4}{\kcalmol} from the TBE reference value, the most accurate result being obtained with M06-2X. 
For the RSH functionals, the automerization barrier is much less sensitive to the amount of longe-range exact exchange.
Another important feature of SF-TD-DFT is the fast convergence of the energy barrier with the size of the basis set. \cite{Loos_2019d} 
With the augmented double-$\zeta$ basis, the SF-TD-DFT results are basically converged to sub-{\kcalmol} accuracy, which is a drastic improvement compared to wave function approaches where this type of convergence is reached with the augmented triple-$\zeta$ basis only.

For the SF-ADC family of methods, the energy differences are much smaller with a maximum deviation of \SI{2}{\kcalmol} between different versions.
In particular, we observe that SF-ADC(2)-s and SF-ADC(3), which respectively scale as $\order*{N^5}$ and $\order*{N^6}$  (where $N$ is the number of basis functions), under- and overestimate the automerization barrier, making SF-ADC(2.5) a good compromise with an error of only \SI{0.18}{\kcalmol} compared to the TBE/aug-cc-pVTZ basis reference value.
Nonetheless, at a $\order*{N^5}$ computational scaling, SF-ADC(2)-s is particularly accurate, even compared to high-order CC methods (see below).
We note that SF-ADC(2)-x [which scales as $\order*{N^6}$] is probably not worth its extra cost [as compared to SF-ADC(2)-s] as it overestimates the energy barrier even more than SF-ADC(3).
This behavior was previously reported by Dreuw's group. \cite{Wormit_2014,Harbach_2014,Dreuw_2015}
Overall, even with the best exchange-correlation functional, SF-TD-DFT is clearly outperformed by the more expensive SF-ADC models.

\alert{We observe that SF-EOM-CCSD/aug-cc-pVTZ tends to underestimate by about \SI{1.5}{\kcalmol} the energy barrier compared to the TBE, an observation in agreement with previous results by Manohar and Krylov. \cite{Manohar_2008}
This can be alleviated by including the triples correction with SF-EOM-CCSD(fT) and SF-EOM-CCSD(dT) (see {\SupInf} where we have reported the data from Ref.~\onlinecite{Manohar_2008}). 
We also note that the SF-EOM-CCSD values for the energy barrier are close to the ones obtained with the more expensive (standard) CC3 method, yet less accurate than values computed with the cheaper SF-ADC(2)-s formalism. 
Note that, in contrast to a previous statement, \cite{Manohar_2008} the (fT) correction performs better than the (dT) correction for the energy barrier. 
However, for the excited states, the situation is reversed (see below).}

Concerning the multi-reference approaches with the minimal (4e,4o) active space, the TBEs are bracketed by the CASPT2 and NEVPT2 values that differ by approximately \SI{1.5}{\kcalmol} for all bases.
In this case, the NEVPT2 values are fairly accurate with differences below half a \si{\kcalmol} compared to the TBEs.
The CASSCF results predict an even lower barrier than CASPT2 due to the well known lack of dynamical correlation at the CASSCF level. 
For the larger (12e,12o) active space, we see larger differences of the order of \SI{3}{\kcalmol} (through all the bases) between CASSCF and the second-order variants (CASPT2 and NEVPT2). 
However, the deviations between CASPT2(12,12) and NEVPT2(12,12) are much smaller than with the minimal active space, with an energy difference of around \SIrange{0.1}{0.2}{\kcalmol} for all bases, CASPT2 being slightly more accurate than NEVPT2 in this case.
For each basis set, both CASPT2(12,12) and NEVPT2(12,12) are less than a \si{\kcalmol} away from the TBEs.
For the two active spaces that we have considered here, the PC- and SC-NEVPT2 schemes provide nearly identical barriers independently of the size of the one-electron basis.

Finally, for the CC family of methods, we observe the usual systematic improvement following the series CCSD $<$ CC3 $<$ CCSDT $<$ CC4 $<$ CCSDTQ, which parallels their increase in computational cost: $\order*{N^6}$, $\order*{N^7}$, $\order*{N^8}$, $\order*{N^9}$, and $\order*{N^{10}}$, respectively.
Note that the introduction of the triple excitations is clearly mandatory to have an accuracy beyond SF-TD-DFT, and we observe that CCSDT is definitely an improvement over its cheaper, approximated version, CC3.

\subsection{Vertical excitation energies}
\label{sec:states}

\subsubsection{{\Dtwo} rectangular geometry}
\label{sec:D2h}

\begin{squeezetable}
\begin{table}
	\caption{
	Spin-flip TD-DFT and ADC vertical excitation energies (with respect to the singlet {\oneAg} ground state) of the {\tBoneg}, {\sBoneg}, and {\twoAg} states of CBD at the {\Dtwo} rectangular equilibrium geometry of the {\oneAg} ground state.}
	\label{tab:sf_D2h}
	\begin{ruledtabular}
		\begin{tabular}{llrrr}
									&	\mc{4}{r}{Excitation energies (eV)}	 \hspace{0.5cm}\\
									\cline{3-5}
			Method		&		Basis	&	{\tBoneg}	&	{\sBoneg}	&	{\twoAg}	\\
			\hline
SF-TD-B3LYP & 6-31+G(d) & $1.706$ & $2.211$ & $3.993$ \\
 & aug-cc-pVDZ & $1.706$ & $2.204$ & $3.992$ \\
 & aug-cc-pVTZ & $1.703$ & $2.199$ & $3.988$ \\
 & aug-cc-pVQZ & $1.703$ & $2.199$ & $3.989$\\[0.1cm]
SF-TD-PBE0 & 6-31+G(d) & $1.687$ & $2.314$ & $4.089$ \\
 & aug-cc-pVDZ & $1.684$ & $2.301$ & $4.085$ \\
 & aug-cc-pVTZ & $1.682$ & $2.296$ & $4.081$ \\
 & aug-cc-pVQZ & $1.682$ & $2.296$ & $4.079$\\[0.1cm]
SF-TD-BH\&HLYP & 6-31+G(d) & $1.552$ & $2.779$ & $4.428$ \\
 & aug-cc-pVDZ & $1.546$ & $2.744$ & $4.422$ \\
 & aug-cc-pVTZ & $1.540$ & $2.732$ & $4.492$ \\
 & aug-cc-pVQZ & $1.540$ & $2.732$ & $4.415$ \\[0.1cm]
 SF-TD-M06-2X & 6-31+G(d) & $1.477$ & $2.835$ & $4.378$ \\
 & aug-cc-pVDZ & $1.467$ & $2.785$ & $4.360$ \\
 & aug-cc-pVTZ & $1.462$ & $2.771$ & $4.357$ \\
 & aug-cc-pVQZ & $1.458$ & $2.771$ & $4.352$ \\[0.1cm]
SF-TD-CAM-B3LYP & 6-31+G(d) & $1.750$ & $2.337$ & $4.140$ \\
 & aug-cc-pVDZ & $1.745$ & $2.323$ & $4.140$ \\
 & aug-cc-pVTZ & $1.742$ & $2.318$ & $4.138$ \\
 & aug-cc-pVQZ & $1.743$ & $2.319$ & $4.138$ \\[0.1cm]
 SF-TD-$\omega$B97X-V & 6-31+G(d) & $1.810$ & $2.377$ & $4.220$ \\
 & aug-cc-pVDZ & $1.800$ & $2.356$ & $4.217$ \\
 & aug-cc-pVTZ & $1.797$ & $2.351$ & $4.213$ \\
 & aug-cc-pVQZ & $1.797$ & $2.351$ & $4.213$ \\[0.1cm]
 SF-TD-LC-$\omega $PBE08 & 6-31+G(d) & $1.917$ & $2.445$ & $4.353$ \\
 & aug-cc-pVDZ & $1.897$ & $2.415$ & $4.346$ \\
 & aug-cc-pVTZ & $1.897$ & $2.415$ & $4.348$ \\
 & aug-cc-pVQZ & $1.897$ & $2.415$ & $4.348$ \\[0.1cm]
SF-TD-M11 & 6-31+G(d) & $1.566$ & $2.687$ & $4.292$ \\
 & aug-cc-pVDZ & $1.546$ & $2.640$ & $4.267$ \\
 & aug-cc-pVTZ & $1.559$ & $2.651$ & $4.300$ \\
 & aug-cc-pVQZ & $1.557$ & $2.650$ & $4.299$ \\[0.1cm]
SF-ADC(2)-s & 6-31+G(d) & $1.577$ & $3.303$ & $4.196$ \\
 & aug-cc-pVDZ & $1.513$ & $3.116$ & $4.114$ \\
 & aug-cc-pVTZ & $1.531$ & $3.099$ & $4.131$ \\
 & aug-cc-pVQZ & $1.544$ & $3.101$ & $4.140$ \\[0.1cm]
SF-ADC(2)-x & 6-31+G(d) & $1.557$ & $3.232$ & $3.728$ \\
 & aug-cc-pVDZ & $1.524$ & $3.039$ & $3.681$ \\
 & aug-cc-pVTZ & $1.539$ & $3.031$ & $3.703$ \\[0.1cm]
SF-ADC(2.5) & 6-31+G(d) & $1.496$ & $3.328$ & $4.219$ \\
 & aug-cc-pVDZ & $1.468$ & $3.148$ & $4.161$ \\
 & aug-cc-pVTZ & $1.475$ & $3.131$ & $4.178$ \\[0.1cm]
SF-ADC(3) & 6-31+G(d) & $1.435$ & $3.352$ & $4.242$ \\
 & aug-cc-pVDZ & $1.422$ & $3.180$ & $4.208$ \\
 & aug-cc-pVTZ & $1.419$ & $3.162$ & $4.224$ \\[0.1cm]
 \alert{SF-EOM-CCSD} & \alert{6-31+G(d)} & \alert{$1.663$} & \alert{$3.515$} & \alert{$4.275$} \\
 & \alert{aug-cc-pVDZ} & \alert{$1.611$} & \alert{$3.315$} & \alert{$4.216$} \\
 & \alert{aug-cc-pVTZ} & \alert{$1.609$} & \alert{$3.293$} & \alert{$4.245$} \\[0.1cm]
		\end{tabular}
	\end{ruledtabular}
\end{table}
\end{squeezetable}

\begin{squeezetable}
\begin{table}
	\caption{
	Vertical excitation energies (with respect to the {\oneAg} ground state) of the {\tBoneg}, {\sBoneg}, and {\twoAg} states of CBD at the {\Dtwo} rectangular equilibrium geometry of the {\oneAg} ground state.	
	The values in square parenthesis have been obtained by extrapolation via the procedure described in the corresponding footnote.
	The TBE/aug-cc-pVTZ values are highlighted in bold.}
	\label{tab:D2h}
	\begin{ruledtabular}
		\begin{tabular}{llrrr}
						&			&	\mc{3}{c}{Excitation energies (eV)}	 \\
									\cline{3-5}
			Method		&		Basis	&	{\tBoneg}	&	{\sBoneg}	&	{\twoAg}	\\
			\hline
CASSCF(4,4) &6-31+G(d)& $1.662$ & $4.657$ & $4.439$ \\
 & aug-cc-pVDZ & $1.672$ & $4.563$ & $4.448$ \\
 & aug-cc-pVTZ & $1.670$ & $4.546$ & $4.441$ \\
 & aug-cc-pVQZ & $1.671$ & $4.549$ & $4.440$ \\[0.1cm]
CASPT2(4,4) &6-31+G(d)& $1.440$ & $3.162$ & $4.115$ \\
 & aug-cc-pVDZ & $1.414$ & $2.971$ & $4.068$ \\
 & aug-cc-pVTZ & $1.412$ & $2.923$ & $4.072$ \\
& aug-cc-pVQZ & $1.417$ & $2.911$ & $4.081$ \\[0.1cm]
SC-NEVPT2(4,4) &6-31+G(d)& $1.407$ & $2.707$ & $4.145$ \\
& aug-cc-pVDZ & $1.381$ & $2.479$ & $4.109$ \\
& aug-cc-pVTZ & $1.379$ & $2.422$ & $4.108$ \\
& aug-cc-pVQZ & $1.384$ & $2.408$ & $4.116$ \\[0.1cm]
PC-NEVPT2(4,4) &6-31+G(d)& $1.409$ & $2.652$ & $4.120$ \\
& aug-cc-pVDZ & $1.384$ & $2.424$ & $4.084$ \\
& aug-cc-pVTZ & $1.382$ & $2.368$ & $4.083$ \\
& aug-cc-pVQZ & $1.387$ & $2.353$ & $4.091$ \\[0.1cm]
CASSCF(12,12) &6-31+G(d)& $1.675$ & $3.924$ & $4.220$ \\
& aug-cc-pVDZ & $1.685$ & $3.856$ & $4.221$ \\
& aug-cc-pVTZ & $1.686$ & $3.844$ & $4.217$ \\
& aug-cc-pVQZ & $1.687$ & $3.846$ & $4.216$ \\[0.1cm]
CASPT2(12,12) &6-31+G(d)& $1.508$ & $3.407$ & $4.099$ \\
& aug-cc-pVDZ & $1.489$ & $3.256$ & $4.044$ \\
& aug-cc-pVTZ & $1.480$ & $3.183$ & $4.043$ \\
& aug-cc-pVQZ & $1.482$ & $3.163$ & $4.047$ \\[0.1cm]
SC-NEVPT2(12,12) &6-31+G(d)& $1.522$ & $3.409$ & $4.130$ \\
& aug-cc-pVDZ & $1.511$ & $3.266$ & $4.093$ \\
& aug-cc-pVTZ & $1.501$ & $3.188$ & $4.086$ \\
& aug-cc-pVQZ & $1.503$ & $3.167$ & $4.088$ \\[0.1cm]
PC-NEVPT2(12,12) &6-31+G(d)& $1.487$ & $3.296$ & $4.103$ \\
& aug-cc-pVDZ & $1.472$ & $3.141$ & $4.064$ \\
& aug-cc-pVTZ & $1.462$ & $3.063$ & $4.056$ \\
& aug-cc-pVQZ & $1.464$ & $3.043$ & $4.059$ \\[0.1cm]
CCSD &6-31+G(d)& $1.346$ & $3.422$ &  \\
 & aug-cc-pVDZ & $1.319$ & $3.226$ &  \\
 & aug-cc-pVTZ & $1.317$ & $3.192$ &  \\
 & aug-cc-pVQZ & $1.323$ & $3.187$& \\[0.1cm]
CC3 &6-31+G(d)& $1.420$ & $3.341$ & $4.658$ \\
 & aug-cc-pVDZ & $1.396$ & $3.158$ & $4.711$ \\
 & aug-cc-pVTZ & $1.402$ & $3.119$ & $4.777$ \\
 & aug-cc-pVQZ & $1.409$ & $3.113$ & $4.774$ \\[0.1cm]
CCSDT &6-31+G(d)& $1.442$ & $3.357$ & $4.311$ \\
 & aug-cc-pVDZ & $1.411$ & $3.175$ & $4.327$ \\
 & aug-cc-pVTZ & $1.411$ & $3.139$ & $4.429$ \\[0.1cm]
CC4 &6-31+G(d)& & $3.343$ & $4.067$ \\
 & aug-cc-pVDZ &  & $3.164$ & $4.040$ \\
 & aug-cc-pVTZ & & $[3.128]$\fnm[1] & $[4.032]$\fnm[2]\\[0.1cm]
CCSDTQ &6-31+G(d)&  $1.464$  & $3.340$ & $4.073$ \\
& aug-cc-pVDZ & $[1.433]$\fnm[3]& $[3.161]$\fnm[4]&  $[4.046]$\fnm[4] \\
& aug-cc-pVTZ & \alert{$[\bf 1.433]$\fnm[5]} & $[\bf 3.125]$\fnm[6]&  $[\bf 4.038]$\fnm[6]\\[0.1cm]
CIPSI &6-31+G(d)& $1.486\pm 0.005$ & $3.348\pm 0.024$ & $4.084\pm 0.012$ \\
& aug-cc-pVDZ & $1.458\pm 0.009$ & $3.187\pm 0.035$ & $4.04\pm 0.04$ \\
& aug-cc-pVTZ & $1.461\pm 0.030$ & $3.142\pm 0.035$ & $4.03\pm 0.09$ \\
			\end{tabular}
	\end{ruledtabular}
	\fnt[1]{Value obtained using CC4/aug-cc-pVDZ corrected by the difference between CCSDT/aug-cc-pVTZ and CCSDT/aug-cc-pVDZ.}
	\fnt[2]{Value obtained using CC4/aug-cc-pVDZ corrected by the difference between PC-NEVPT2(12,12)/aug-cc-pVTZ and PC-NEVPT2(12,12)/aug-cc-pVDZ.}
	\fnt[3]{Value obtained using CCSDTQ/6-31+G(d) corrected by the difference between CCSDT/aug-cc-pVDZ and CCSDT/6-31+G(d).}
	\fnt[4]{Value obtained using CCSDTQ/6-31+G(d) corrected by the difference between CC4/aug-cc-pVDZ and CC4/6-31+G(d).}
	\fnt[5]{Value obtained using CCSDTQ/aug-cc-pVDZ corrected by the difference between CCSDT/aug-cc-pVTZ and CCSDT/aug-cc-pVDZ.}
	\fnt[6]{TBE value obtained using CCSDTQ/aug-cc-pVDZ corrected by the difference between CC4/aug-cc-pVTZ and CC4/aug-cc-pVDZ.}
\end{table}
\end{squeezetable}

\begin{figure*}
	\includegraphics[width=0.8\linewidth]{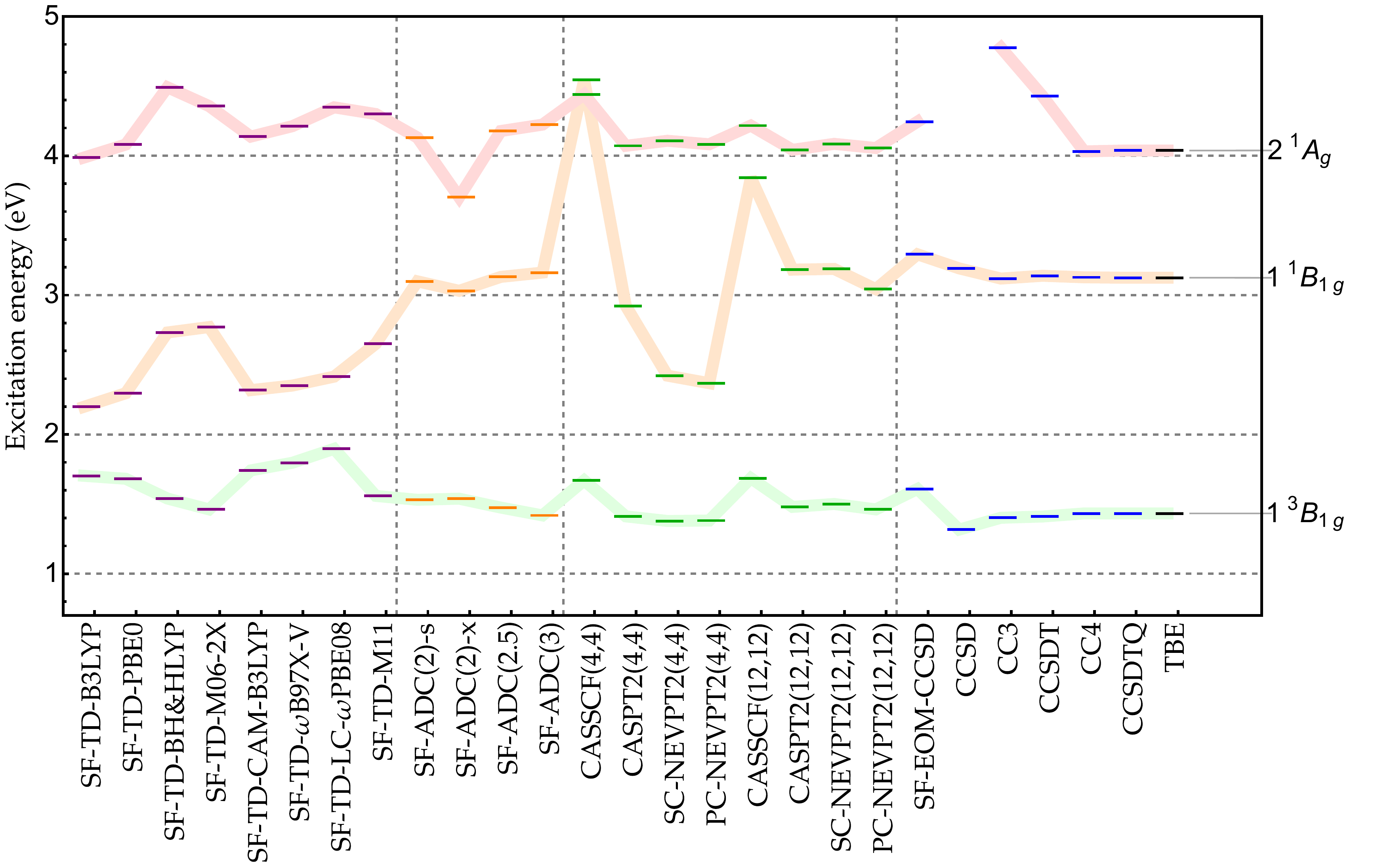}
	\caption{Vertical excitation energies of the {\tBoneg}, {\sBoneg}, and {\twoAg} states at the {\Dtwo} rectangular equilibrium geometry of the {\oneAg} ground state using the aug-cc-pVTZ basis.
	See {\SupInf} for the raw data.}
\label{fig:D2h}
\end{figure*}

Table \ref{tab:sf_D2h} reports, at the {\Dtwo} rectangular equilibrium geometry of the {\oneAg} ground state, the vertical transition energies associated with the {\tBoneg}, {\sBoneg}, and {\twoAg} states obtained using the spin-flip formalism, while Table \ref{tab:D2h} gathers the same quantities obtained with the multi-reference, CC, and CIPSI methods. 
Considering the aug-cc-pVTZ basis, the evolution of the vertical excitation energies with respect to the level of theory is illustrated in Fig.~\ref{fig:D2h}. 

At the CC3/aug-cc-pVTZ level, the percentage of single excitation involved in the {\tBoneg}, {\sBoneg}, and {\twoAg} are 99\%, 95\%, and 1\%, respectively.
Therefore, the two formers are dominated by single excitations, while the latter state corresponds to a genuine double excitation.

First, let us discuss basis set effects at the SF-TD-DFT level (Table \ref{tab:sf_D2h}).
As expected, these are found to be small and the results are basically converged to the complete basis set limit with the triple-$\zeta$ basis, which is definitely not the case for the wave function methods. \cite{Giner_2019}
Regarding now the accuracy of the vertical excitation energies, again, we see that, for {\tBoneg} and {\sBoneg}, the functionals with the largest amount of short-range exact exchange (\eg, BH\&HLYP, M06-2X, and M11) are the most accurate.
Functionals with a large share of exact exchange are known to perform best in the SF-TD-DFT framework as the Hartree-Fock exchange term is the only non-vanishing term in the spin-flip block. \cite{Shao_2003}
However, their overall accuracy remains average especially for the singlet states, {\sBoneg} and {\twoAg}, with error of the order of \SIrange{0.2}{0.5}{\eV} compared to the TBEs.
The triplet state, {\tBoneg}, is much better described with errors below \SI{0.1}{\eV}. 
Surprisingly, for the doubly-excited state, {\twoAg}, the hybrid functionals with a low percentage of exact exchange (B3LYP and PBE0) are the best performers with absolute errors below \SI{0.05}{\eV}.
Note that, as evidenced by the data reported in {\SupInf}, none of these states exhibit a strong spin contamination.

Second, we discuss the various SF-ADC schemes (Table \ref{tab:sf_D2h}), \ie, SF-ADC(2)-s, SF-ADC(2)-x, and SF-ADC(3). 
At the SF-ADC(2)-s level, going from the smallest 6-31+G(d) basis to the largest aug-cc-pVQZ basis induces a small decrease in vertical excitation energies of  \SI{0.03}{\eV} (\SI{0.06}{\eV}) for the {\tBoneg} ({\twoAg}) state, while the transition energy of the {\sBoneg} state drops more significantly by about \SI{0.2}{\eV}.
[The SF-ADC(2)-x and SF-ADC(3) calculations with aug-cc-pVQZ were not feasible with our computational resources.]
These basis set effects are fairly transferable to the other wave function methods that we have considered here. 
This further motivates the ``pyramidal'' extrapolation scheme that we have employed to produce the TBE values (see Sec.~\ref{sec:TBE}).
Again, the extended version, SF-ADC(2)-x, does not seem to be relevant in the present context with much larger errors than the other schemes.
Also, as reported previously, \cite{Loos_2020d} SF-ADC(2)-s and SF-ADC(3) have mirror error patterns making SF-ADC(2.5) particularly accurate except for the doubly-excited state {\twoAg} where the error with respect to the TBE (\SI{0.140}{\eV}) is larger than the SF-ADC(2)-s error (\SI{0.093}{\eV}). 

\alert{Interestingly, we observe that the SF-EOM-CCSD excitation energies are systematically larger than the TBEs by approximately \SI{0.2}{\eV} with a nice consistency throughout the various (singly- and doubly-) excited states. 
Moreover, SF-EOM-CCSD excitation energies are somehow closer to their SF-ADC(2)-s analogs (with an energy difference of about \SI{0.1}{\eV}) than the other schemes as already noticed by LeFrançois and co-workers. \cite{Lefrancois_2015}
We see that the SF-EOM-CCSD excitations energies for the triplet state are larger of about \SI{0.3}{\eV} compared to the CCSD ones, which was also pointed out in the study of Manohar and Krylov. \cite{Manohar_2008}
Again, our SF-EOM-CCSD results are very similar to the ones obtained in previous studies \cite{Manohar_2008,Lefrancois_2015}. 
We can logically expect similar trend for SF-EOM-CCSD(fT) and SF-EOM-CCSD(dT) that lower the excitation energies and tend to be in better agreement with respect to the TBE (see {\SupInf}). 
Note that the (dT) correction slightly outperforms the (fT) correction as previously observed \cite{Manohar_2008} and theoretically expected.}

Let us now move to the discussion of the results obtained with standard wave function methods that are reported in Table \ref{tab:D2h}.
Regarding the \alert{multi-reference} calculations, the most striking result is the poor description of the {\sBoneg} ionic state, especially with the (4e,4o) active space where CASSCF predicts this state higher in energy than the {\twoAg} state. 
Of course, the PT2 correction is able to correct the state ordering problem but cannot provide quantitative excitation energies due to the poor zeroth-order treatment.
Another ripple effect of the unreliability of the reference wave function is the large difference between CASPT2 and NEVPT2 that differ by half an \si{\eV}. 
This feature is characteristic of the inadequacy of the active space to model such a state.
\alert{Additional MRCI and MRCI+Q calculations (reported in the {\SupInf}) confirm this.}
For the two other states, {\tBoneg} and {\twoAg}, the errors at the CASPT2(4,4) and NEVPT2(4,4) levels are much smaller (below \SI{0.1}{\eV}).
Using a larger active space resolves most of these issues: CASSCF predicts the correct state ordering (though the ionic state is still badly described in term of energetics), CASPT2 and NEVPT2 excitation energies are much closer, and their accuracy is often improved (especially for the triplet and doubly-excited states) although it is difficult to reach chemical accuracy (\ie, an error below \SI{0.043}{\eV}) on a systematic basis.

Finally, for the CC models (Table \ref{tab:D2h}), the two states with a large $\%T_1$ value, {\tBoneg} and {\sBoneg}, are already extremely accurate at the CC3 level, and systematically improved by CCSDT and CC4.
This trend is in line with the observations made on the QUEST database. \cite{Veril_2021}
For the doubly-excited state, {\twoAg}, the convergence of the CC expansion is much slower but it is worth pointing out that the inclusion of approximate quadruples via CC4 is particularly effective, as observed in an earlier work. \cite{Loos_2021}
The CCSDTQ excitation energies (which are used to define the TBEs) are systematically within the error bar of the CIPSI extrapolations, which confirms the outstanding performance of CC methods that include quadruple excitations in the context of excited states.

\subsubsection{{\Dfour} square-planar geometry}
\label{sec:D4h}

\begin{squeezetable}
\begin{table}
	\caption{
	Spin-flip TD-DFT and ADC vertical excitation energies (with respect to the singlet {\sBoneg} ground state) of the {\Atwog}, {\Aoneg}, and {\Btwog} states of CBD at the {\Dfour} square-planar equilibrium geometry of the {\Atwog} state.}
	\label{tab:sf_D4h}
	\begin{ruledtabular}
		\begin{tabular}{llrrr}
									&	\mc{4}{r}{Excitation energies (eV)}	 \hspace{0.5cm}\\
									\cline{3-5}
			Method	& Basis				&	{\Atwog}	&	 {\Aoneg}	&	 {\Btwog}	\\
			\hline
SF-TD-B3LYP & 6-31+G(d) & $-0.016$ & $0.487$ & $0.542$ \\
 & aug-cc-pVDZ & $-0.019$ & $0.477$ & $0.536$ \\
 & aug-cc-pVTZ & $-0.020$ & $0.472$ & $0.533$ \\
 & aug-cc-pVQZ & $-0.020$ & $0.473$ & $0.533$ \\[0.1cm]
SF-TD-PBE0 & 6-31+G(d) & $-0.012$ & $0.618$ & $0.689$ \\
 & aug-cc-pVDZ & $-0.016$ & $0.602$ & $0.680$ \\
 & aug-cc-pVTZ & $-0.019$ & $0.597$ & $0.677$ \\
 & aug-cc-pVQZ & $-0.018$ & $0.597$ & $0.677$ \\[0.1cm]
SF-TD-BH\&HLYP & 6-31+G(d) & $0.064$ & $1.305$ & $1.458$ \\
 & aug-cc-pVDZ & $0.051$ & $1.260$ & $1.437$ \\
 & aug-cc-pVTZ & $0.045$ & $1.249$ & $1.431$ \\
 & aug-cc-pVQZ & $0.046$ & $1.250$ & $1.432$ \\[0.1cm]
SF-TD-M06-2X & 6-31+G(d) & $0.102$ & $1.476$ & $1.640$ \\
 & aug-cc-pVDZ & $0.086$ & $1.419$ & $1.611$ \\
 & aug-cc-pVTZ & $0.078$ & $1.403$ & $1.602$ \\
 & aug-cc-pVQZ & $0.079$ & $1.408$ & $1.607$ \\[0.1cm]
SF-TD-CAM-B3LYP & 6-31+G(d) & $0.021$ & $0.603$ & $0.672$ \\
 & aug-cc-pVDZ & $0.012$ & $0.585$ & $0.666$ \\
 & aug-cc-pVTZ & $0.010$ & $0.580$ & $0.664$ \\
 & aug-cc-pVQZ & $0.010$ & $0.580$ & $0.664$ \\[0.1cm]
SF-TD-$\omega $B97X-V & 6-31+G(d) & $0.040$ & $0.600$ & $0.670$ \\
 & aug-cc-pVDZ & $0.029$ & $0.576$ & $0.664$ \\
 & aug-cc-pVTZ & $0.026$ & $0.572$ & $0.662$ \\
 & aug-cc-pVQZ & $0.026$ & $0.572$ & $0.662$ \\[0.1cm]
 SF-TD-LC-$\omega $PBE08 & 6-31+G(d) & $0.078$ & $0.593$ & $0.663$ \\
 & aug-cc-pVDZ & $0.060$ & $0.563$ & $0.659$ \\
 & aug-cc-pVTZ & $0.058$ & $0.561$ & $0.658$ \\
 & aug-cc-pVQZ & $0.058$ & $0.561$ & $0.659$ \\[0.1cm]
SF-TD-M11 & 6-31+G(d) & $0.102$ & $1.236$ & $1.374$ \\
 & aug-cc-pVDZ & $0.087$ & $1.196$ & $1.362$ \\
 & aug-cc-pVTZ & $0.081$ & $1.188$ & $1.359$ \\
 & aug-cc-pVQZ & $0.080$ & $1.185$ & $1.357$ \\[0.1cm]
SF-ADC(2)-s & 6-31+G(d) & $0.345$ & $1.760$ & $2.096$ \\
 & aug-cc-pVDZ & $0.269$ & $1.656$ & $1.894$ \\
 & aug-cc-pVTZ & $0.256$ & $1.612$ & $1.844$ \\[0.1cm]
SF-ADC(2)-x & 6-31+G(d) & $0.264$ & $1.181$ & $1.972$ \\
 & aug-cc-pVDZ & $0.216$ & $1.107$ & $1.760$ \\
 & aug-cc-pVTZ & $0.212$ & $1.091$ & $1.731$ \\[0.1cm]
SF-ADC(2.5) & 6-31+G(d) & $0.234$ & $1.705$ & $2.087$ \\
 & aug-cc-pVDZ & $0.179$ & $1.614$ & $1.886$ \\
 & aug-cc-pVTZ & $0.168$ & $1.594$ & $1.849$ \\[0.1cm]
SF-ADC(3) & 6-31+G(d) & $0.123$ & $1.650$ & $2.078$ \\
 & aug-cc-pVDZ & $0.088$ & $1.571$ & $1.878$ \\
 & aug-cc-pVTZ & $0.079$ & $1.575$ & $1.853$ \\[0.1cm]
\alert{SF-EOM-CCSD} & \alert{6-31+G(d)} & \alert{$0.446$} & \alert{$1.875$} & \alert{$2.326$} \\
& \alert{aug-cc-pVDZ} & \alert{$0.375$} & \alert{$1.776$} & \alert{$2.102$} \\
& \alert{aug-cc-pVTZ}& \alert{$0.354$} & \alert{$1.768$} & \alert{$2.060$} \\
		\end{tabular}
	\end{ruledtabular}
	
\end{table}
\end{squeezetable}

\begin{squeezetable}
\begin{table}
	\caption{
	Vertical excitation energies (with respect to the {\sBoneg} ground state) of the {\Atwog}, {\Aoneg}, and {\Btwog} states of CBD at the {\Dfour} square-planar equilibrium geometry of the {\Atwog} state.
	The values in square brackets have been obtained by extrapolation via the procedure described in the corresponding footnote.
	The TBE/aug-cc-pVTZ values are highlighted in bold.}
	\label{tab:D4h}
	\begin{ruledtabular}
		\begin{tabular}{llrrr}
									&	\mc{3}{r}{Excitation energies (eV)}	 \hspace{0.1cm}\\
									\cline{3-5}
			Method	& Basis				&	{\Atwog}	&	 {\Aoneg}	&	{\Btwog}	\\
			\hline
CASSCF(4,4) & 6-31+G(d) & $0.447$ & $2.257$ & $3.549$ \\
 & aug-cc-pVDZ & $0.438$ & $2.240$ & $3.443$ \\
 & aug-cc-pVTZ & $0.434$ & $2.234$ & $3.424$ \\
 & aug-cc-pVQZ & $0.435$ & $2.235$ & $3.427$ \\[0.1cm]
CASPT2(4,4) & 6-31+G(d) & $0.176$ & $1.588$ & $1.899$ \\
 & aug-cc-pVDZ & $0.137$ & $1.540$ & $1.708$ \\
 & aug-cc-pVTZ & $0.128$ & $1.506$ & $1.635$ \\
 & aug-cc-pVQZ & $0.128$ & $1.498$ & $1.612$ \\[0.1cm]
SC-NEVPT2(4,4) & 6-31+G(d) & $0.083$ & $1.520$ & $1.380$ \\
 & aug-cc-pVDZ & $0.037$ & $1.465$ & $1.140$ \\
 & aug-cc-pVTZ & $0.024$ & $1.428$ & $1.055$ \\
 & aug-cc-pVQZ & $0.024$ & $1.420$ & $1.030$ \\[0.1cm]
PC-NEVPT2(4,4) & 6-31+G(d) & $0.085$ & $1.496$ & $1.329$ \\
 & aug-cc-pVDZ & $0.039$ & $1.440$ & $1.088$ \\
 & aug-cc-pVTZ & $0.026$ & $1.403$ & $1.003$ \\
 & aug-cc-pVQZ & $0.026$ & $1.395$ & $0.977$ \\[0.1cm]
CASSCF(12,12) & 6-31+G(d) & $0.386$ & $1.974$ & $2.736$ \\
 & aug-cc-pVDZ & $0.374$ & $1.947$ & $2.649$ \\
 & aug-cc-pVTZ & $0.370$ & $1.943$ & $2.634$ \\
 & aug-cc-pVQZ & $0.371$ & $1.945$ & $2.637$ \\[0.1cm]
CASPT2(12,12) & 6-31+G(d) & $0.235$ & $1.635$ & $2.170$ \\
 & aug-cc-pVDZ & $0.203$ & $1.588$ & $2.015$ \\
 & aug-cc-pVTZ & $0.183$ & $1.538$ & $1.926$ \\
 & aug-cc-pVQZ & $0.179$ & $1.522$ & $1.898$ \\[0.1cm]
SC-NEVPT2(12,12) & 6-31+G(d) & $0.218$ & $1.644$ & $2.143$ \\
 & aug-cc-pVDZ & $0.189$ & $1.600$ & $1.991$ \\
 & aug-cc-pVTZ & $0.165$ & $1.546$ & $1.892$ \\
 & aug-cc-pVQZ & $0.160$ & $1.529$ & $1.862$ \\[0.1cm]
PC-NEVPT2(12,12) & 6-31+G(d) & $0.189$ & $1.579$ & $2.020$ \\
 & aug-cc-pVDZ & $0.156$ & $1.530$ & $1.854$ \\
 & aug-cc-pVTZ & $0.131$ & $1.476$ & $1.756$ \\
 & aug-cc-pVQZ & $0.126$ & $1.460$ & $1.727$ \\[0.1cm]
CCSD & 6-31+G(d) & $0.148$ & $1.788$ &  \\
 & aug-cc-pVDZ & $0.100$ & $1.650$ &  \\
 & aug-cc-pVTZ & $0.085$ & $1.600$ &  \\
 & aug-cc-pVQZ & $0.084$ & $1.588$ &  \\[0.1cm]
CC3 & 6-31+G(d) &  & $1.809$ & $2.836$ \\
 & aug-cc-pVDZ &  & $1.695$ & $2.646$ \\
 & aug-cc-pVTZ &  & $1.662$ & $2.720$ \\[0.1cm]
CCSDT & 6-31+G(d) & $0.210$ & $1.751$ & $2.565$ \\
 & aug-cc-pVDZ & $0.165$ & $1.659$ & $2.450$ \\
 & aug-cc-pVTZ & $0.149$ & $1.631$ & $2.537$ \\[0.1cm]
CC4 & 6-31+G(d) &  & $1.604$ & $2.121$ \\
 & aug-cc-pVDZ &  & $1.539$ & $1.934$ \\
 & aug-cc-pVTZ &  & $[1.511 ]$\fnm[1] &$[1.836 ]$\fnm[2]  \\[0.1cm]
CCSDTQ & 6-31+G(d) & $0.205$ & $1.593$ & $2.134$ \\
& aug-cc-pVDZ & $[0.160]$\fnm[3] & $[1.528 ]$\fnm[4]&$[1.947]$\fnm[4] \\
& aug-cc-pVTZ & $[\bf 0.144]$\fnm[5]  & $[\bf 1.500 ]$\fnm[6]&$[\bf 1.849]$\fnm[6] \\	[0.1cm]
CIPSI & 6-31+G(d) & $0.201\pm 0.003$ & $1.602\pm 0.007$ & $2.13\pm 0.04$ \\
 & aug-cc-pVDZ & $0.157\pm 0.003$ & $1.587\pm 0.005$ & $2.102\pm 0.027$ \\
 & aug-cc-pVTZ & $0.17\pm 0.03$ & $1.63\pm 0.05$ &  \\
		\end{tabular}
	\end{ruledtabular}
	\fnt[1]{Value obtained using CC4/aug-cc-pVDZ corrected by the difference between CCSDT/aug-cc-pVTZ and CCSDT/aug-cc-pVDZ.}
	\fnt[2]{Value obtained using CC4/aug-cc-pVDZ corrected by the difference between PC-NEVPT2(12,12)/aug-cc-pVTZ and PC-NEVPT2(12,12)/aug-cc-pVDZ.}
	\fnt[3]{Value obtained using CCSDTQ/6-31+G(d) corrected by the difference between CCSDT/aug-cc-pVDZ and CCSDT/6-31+G(d).}
	\fnt[4]{Value obtained using CCSDTQ/6-31+G(d) corrected by the difference between CC4/aug-cc-pVDZ and CC4/6-31+G(d).}
	\fnt[5]{TBE value obtained using CCSDTQ/aug-cc-pVDZ corrected by the difference between CCSDT/aug-cc-pVTZ and CCSDT/aug-cc-pVDZ.}
	\fnt[6]{TBE value obtained using CCSDTQ/aug-cc-pVDZ corrected by the difference between CC4/aug-cc-pVTZ and CC4/aug-cc-pVDZ.}
\end{table}
\end{squeezetable}

\begin{figure*}
	\includegraphics[width=0.8\linewidth]{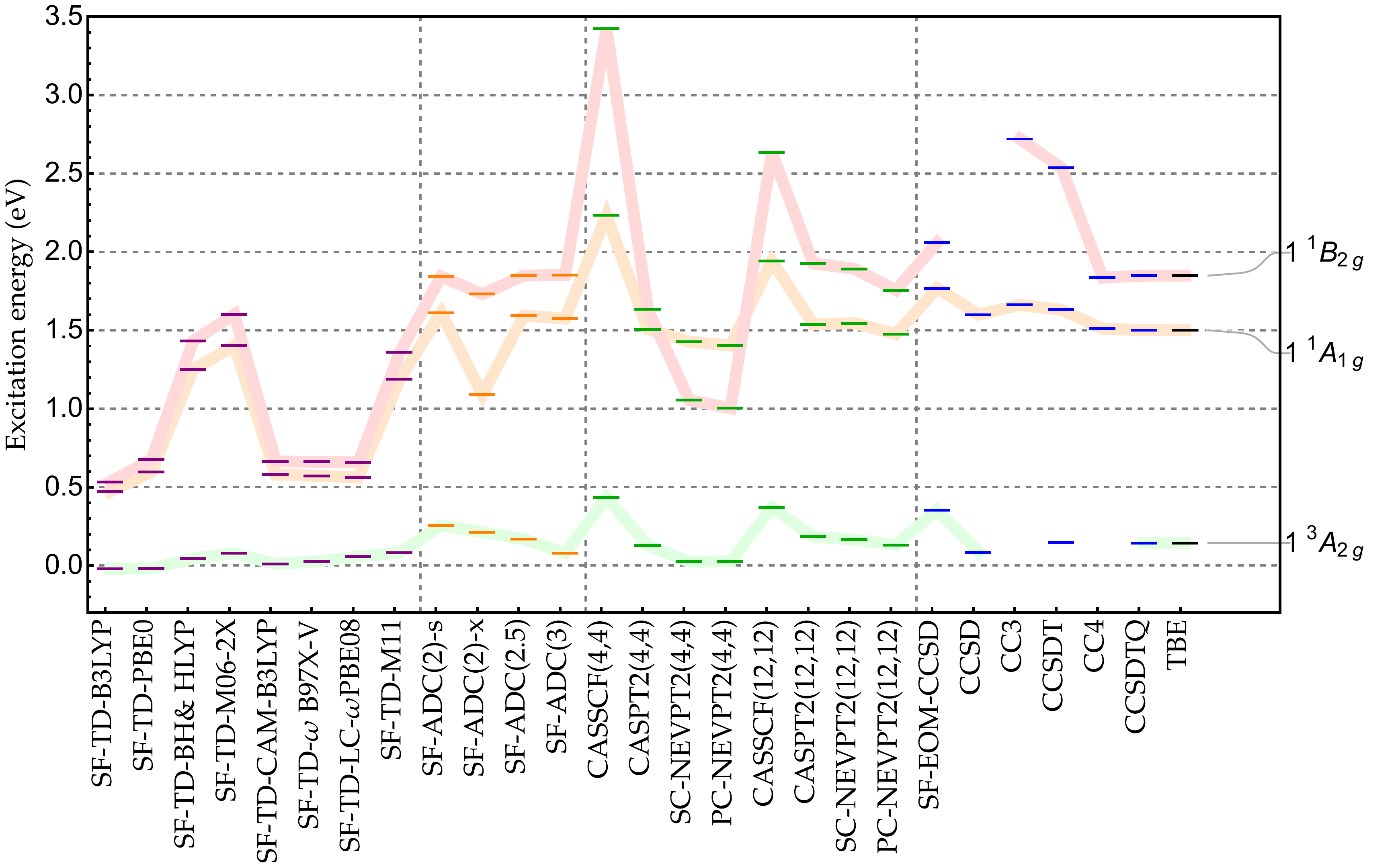}
	\caption{Vertical excitation energies (in \si{\eV}) of the {\Atwog}, {\Aoneg}, and {\Btwog} states at the {\Dfour} square-planar equilibrium geometry of the {\Atwog} state using the aug-cc-pVTZ basis.
	See {\SupInf} for the raw data.} 
	\label{fig:D4h}
\end{figure*}

In Table \ref{tab:sf_D4h}, we report, at the {\Dfour} square planar equilibrium geometry of the {\Atwog} state, the vertical transition energies associated with the {\Atwog}, {\Aoneg}, and {\Btwog} states obtained using the spin-flip formalism, while Table \ref{tab:D4h} gathers the same quantities obtained with the multi-reference, CC, and CIPSI methods. 
The vertical excitation energies computed at various levels of theory are depicted in Fig.~\ref{fig:D4h} for the aug-cc-pVTZ basis.
Unfortunately, due to technical limitations, we could not compute $\%T_1$ values associated with the {\Atwog}, {\Aoneg}, and {\Btwog} excited states in the {\Dfour} symmetry.
However, it is clear from the inspection of the wave function that, with respect to the {\sBoneg} ground state, {\Atwog} and {\Btwog} are dominated by single excitations, while {\Aoneg} has a strong double excitation character.

As for the previous geometry we start by discussing the SF-TD-DFT results (Table \ref{tab:sf_D4h}), and in particular the singlet-triplet gap, \ie, the energy difference between {\sBoneg} and {\Atwog}.
For all functionals, this gap is small (basically below \SI{0.1}{\eV} while the TBE value is \SI{0.144}{\eV}) but it is worth mentioning that B3LYP and PBE0 incorrectly deliver a negative singlet-triplet gap (hence a triplet ground state at this geometry).
Increasing the fraction of exact exchange in hybrids or relying on RSHs (even with a small amount of short-range exact exchange) allows to recover a positive gap and a singlet ground state.
At the SF-TD-DFT level, the energy gap between the two singlet excited states, {\Aoneg} and {\Btwog}, is particularly small and grows moderately with the amount of exact exchange at short range.
The influence of the exact exchange on the singlet energies is quite significant with an energy difference of the order of \SI{1}{\eV} between the functional with the smallest amount of exact exchange (B3LYP) and the functional with the largest amount (M06-2X).
As for the excitation energies computed on the {\Dtwo} ground-state equilibrium structure and the automerization barrier, the functionals with a large fraction of short-range exact exchange yield more accurate results.
Yet, the transition energy to {\Btwog} is off by half an \si{\eV} compared to the TBE for BH\&HLYP and M11, while the doubly-excited state is much closer to the reference value (errors of \SI{-0.251}{} and \SI{-0.312}{\eV} for BH\&HLYP and M11, respectively).
With errors of \SI{-0.066}{}, \SI{-0.097}{}, and \SI{-0.247}{\eV} for {\Atwog}, {\Aoneg}, and {\Btwog}, M06-2X is the best performer here. 
Again, for all the excited states, the basis set effects are extremely small at the SF-TD-DFT level.
We emphasize that the $\expval*{S^2}$ values reported in {\SupInf} indicate again that there is no significant spin contamination in these excited states.

Next, we discuss the various ADC schemes (Table \ref{tab:sf_D4h}).
Globally, we observe similar trends as those noted in Sec.~\ref{sec:D2h}.
Concerning the singlet-triplet gap, each scheme predicts it to be positive.
Although it provides a decent singlet-triplet gap value, SF-ADC(2)-x seems to particularly struggle with the singlet excited states ({\Aoneg} and {\Btwog}), especially for the doubly-excited state {\Aoneg} where it underestimates the vertical excitation energy by \SI{0.4}{\eV}. 
Again, averaging the SF-ADC(2)-s and SF-ADC(3) transition energies is beneficial in most cases at the exception of {\Aoneg}.
Although the basis set effects are larger than at the SF-TD-DFT level, they remain quite moderate at the SF-ADC level, and this holds for wave function methods in general.
\alert{Concerning the SF-EOM-CCSD excitation energies at the {\Dfour} square planar equilibrium geometry, very similar conclusions to the ones provided in the previous section dealing with the excitation energies at the {\Dtwo} rectangular equilibrium geometry can be drawn: (i) SF-EOM-CCSD systematically and consistently overestimates the TBEs by approximately \SI{0.2}{\eV} and is less accurate than SF-ADC(2)-s, (ii) the non-iterative triples corrections tend to give a better agreement with respect to the TBE (see {\SupInf}), and (iii) the (dT) correction performs better than the (fT) one.}

Let us turn to the multi-reference results (Table \ref{tab:D4h}).
For both active spaces, expectedly, CASSCF does not provide a quantitive energetic description, although it is worth mentioning that the right state ordering is preserved.
This is, of course, magnified with the (4e,4o) active space for which the second-order perturbative treatment is unable to provide a satisfying description due to the limited active space.
In particular SC-NEVPT2(4,4)/aug-cc-pVTZ and PC-NEVPT2(4,4)/aug-cc-pVTZ underestimate the singlet-triplet gap by \SI{0.072}{} and \SI{0.097}{\eV} and, more importantly, flip the ordering of {\Aoneg} and {\Btwog}.
Although {\Aoneg} is not badly described, the excitation energy of the ionic state {\Btwog} is off by almost \SI{1}{\eV}.
Thanks to the IPEA shift in CASPT2(4,4), the singlet-triplet gap is accurate and the state ordering remains correct but the ionic state is still far from being well described.
The (12e,12o) active space significantly alleviates these effects, and, as usual now, the agreement between CASPT2 and NEVPT2 is very much improved for each state, though the accuracy of \alert{multi-reference} approaches remains questionable for the ionic state with, \emph{e.g.,} an error up to \SI{-0.093}{\eV} at the PC-NEVPT2(12,12)/aug-cc-pVTZ level.

Finally, let us analyze the excitation energies computed with various CC models that are gathered in Table \ref{tab:D4h}.
As mentioned in Sec.~\ref{sec:CC}, we remind the reader that these calculations are performed by considering the {\Aoneg} state as reference, and that, therefore, 
{\sBoneg} and {\Btwog} are obtained as a de-excitation and an excitation, respectively. 
Consequently, with respect to {\Aoneg}, {\sBoneg} has a dominant double excitation character, while {\Btwog} have a dominant single excitation character. 
This explains why one observes a slower convergence of the transition energies in the case of {\sBoneg} as shown in Fig.~\ref{fig:D4h}.
It is clear from the results of Table \ref{tab:D4h} that, if one wants to reach high accuracy with such a computational strategy, it is mandatory to include quadruple excitations.
Indeed, at the CCSDT/aug-cc-pVTZ level, the singlet-triplet gap is already very accurate (off by \SI{0.005}{\eV} only) while the excitation energies of the singlet states are still \SI{0.131}{} and \SI{0.688}{\eV} away from their respective TBE.
These deviations drop to \SI{0.011}{} and \SI{-0.013}{\eV} at the CC4/aug-cc-pVTZ level.
As a final comment, we can note that the CCSDTQ-based TBEs and the CIPSI results are consistent if one takes into account the extrapolation error (see Sec.~\ref{sec:SCI}).

\section{Conclusions}
\label{sec:conclusion}

In the present study, we have benchmarked a larger number of computational methods on the automerization barrier and the vertical excitation energies of cyclobutadiene in its square ({\Dfour}) and rectangular ({\Dtwo}) geometries, for which we have defined theoretical best estimates based on extrapolated CCSDTQ/aug-cc-pVTZ data.

The main take-home messages of the present work can be summarized as follows:
\begin{itemize}

\item Within the SF-TD-DFT framework, we advice to use exchange-correlation (hybrids or range-separated hybrids) with a large fraction of short-range exact exchange. 
This has been shown to be clearly beneficial for the automerization barrier and the vertical excitation energies computed on both the {\Dtwo} and {\Dfour} equilibrium geometries.

\item At the SF-ADC level, we have found that, as expected, the extended scheme, SF-ADC(2)-x, systematically worsen the results compared to the cheaper standard version, SF-ADC(2)-s. 
Moreover, as previously reported, SF-ADC(2)-s and SF-ADC(3) have opposite error patterns which means that SF-ADC(2.5) emerges as an excellent compromise.

\item \alert{SF-EOM-CCSD shows similar performance as the cheaper SF-ADC(2)-s formalism, especially for the excitation energies. 
As previously reported, the two variants including non-iterative triples corrections, SF-EOM-CCSD(dT) and SF-EOM-CCSD(fT), improve the results, the (dT) correction performing slightly better for the vertical excitation energies computed at the {\Dtwo} and {\Dfour} equilibrium geometries.}

\item For the {\Dfour} square planar structure, a faithful energetic description of the excited states is harder to reach at the SF-TD-DFT level because of the strong multi-configurational character.
In such scenario, the SF-TD-DFT excitation energies can exhibit errors of the order of \SI{1}{\eV} compared to the TBEs. 
However, it was satisfying to see that the spin-flip version of ADC can lower these errors to \SIrange{0.1}{0.2}{\eV}.

\item Concerning the \alert{multi-reference} methods, we have found that while NEVPT2 and CASPT2 can provide different excitation energies for the small (4e,4o) active space, the results become highly similar when the larger (12e,12o) active space is considered. 
From a more general perspective, a significant difference between NEVPT2 and CASPT2 is usually not a good omen and can be seen as a clear warning sign that the active space is too small or poorly chosen.
The ionic states remain a struggle for both CASPT2 and NEVPT2, even with the (12e,12o) active space.

\item In the context of CC methods, although the inclusion of triple excitations (via CC3 or CCSDT) yields very satisfactory results in most cases, the inclusion of quadruples excitation (via CC4 or CCSDTQ) is mandatory to reach high accuracy (especially in the case of doubly-excited states).
Finally, we point out that, considering the error bar related to the CIPSI extrapolation procedure, CCSDTQ and CIPSI yield equivalent excitation energies, hence confirming the outstanding accuracy of CCSDTQ in the context of molecular excited states.

\end{itemize}

\acknowledgements{
EM, AS, and PFL acknowledge funding from the European Research Council (ERC) under the European Union's Horizon 2020 research and innovation programme (Grant agreement No.~863481).
This work used the HPC ressources from CALMIP (Toulouse) under allocation 2022-18005 and from the CCIPL center (Nantes).

\section*{Supporting Information Available}
Included in the {\SupInf} are the raw data, additional calculations and geometries, and the Cartesian coordinates of the various optimized geometries.

\bibliography{CBD}
\end{document}